\documentclass[times, 12pt]{article}
\usepackage{mystyle}
\usepackage{law}

\newtheorem{definition}{Definition}

\newlength{\lineheight}
\newcounter{saved}
\def\save{\saveitem{saved}\addtocounter{saved}{-1}}

\def\resume{\stepcounter{saved}\setitem{saved}}
{\begin{rules} \resume}%
{\save \end{rules}}

\title{Discriminating Defense Against  DDoS Attacks\\a Novel Approach}
\author{Naftaly H. Minsky\\
Rutgers University  Emeritus Professor\\
Email: \TT{nminsky@gmail.com } }

\begin{document}

\maketitle      
\tableofcontents
\bibliographystyle{plain}
\newpage

\begin{abstract}
A recent  paper (circa 2020) by Osterwile et al., entitled
\emph{``21 Years of Distributed Denial of Service: A Call to Action''},
states: \emph{``We are falling behind in the war against
distributed denial-of-service attacks. Unless
we act now, the future of the Internet could be
at stake.''}

And an earlier (circa 2007) paper by Peng et al. states: \emph{``a key challenge for the
defense [against DDoS attacks] is how to discriminate legitimate
requests for service from malicious access attempts.''}  This
challenge has not been met yet, which is, arguably, a major reason for
the dire situation described by Osterwile et al.---thirteen years
later.

This paper attempts to meet an approximation to this challenge, by enabling a site to
define the kind of messages that it consider \emph{important}, and
introducing  an unambiguous  criterion
of discrimination between messages that a given site defined as
important to it, 
and all other messages sent to it.

 And this paper introduces two anti-DDoS
mechanisms based on this criterion, which ensures the delivery to a
given site of practically all the important messages sent to it, even
when it is under attack.
One of these mechanism relies on lightweight support by routers, and the
other one  does not.

\end{abstract}

\s{Introduction}\label{intro}
A circa 2007  paper \cite{pen07-1} by Peng et al. states:
\emph{``a key challenge for the
defense [against DDoS attacks] is how to discriminate legitimate requests for
service from malicious access attempts.''}
A similar view was expressed even earlier  (circa 2002) by
 Ioannidis and Bellovin in their famous pushback paper \cite{ioa02-1},
 as follows: \emph{``If we could unequivocally detect packets belonging to an
attack and drop just those, the problem would be
solved.''}

 This challenge has not been met yet, which is, arguably, a major
reason for the dire situation described  by Osterwile et al.
in their   (circa 2020) paper \cite{ost20-1}, entitled
\emph{``21 Years of Distributed Denial of Service: A Call to
Action''}. This paper
states: \emph{``We are falling behind in the war against
distributed denial-of-service attacks.''} 
The seriousness of the situation is reflected by the large
number of   scholarly papers published on this topic recently. In particular,
Google  scholar reported on 2/21/21, that there were 4184 papers
 published during the past five years with the term ``DDoS'' or the phrase
 ``Distributed Denial of Service'' in their title.

 \p{The Goal of this Paper is Twofold:}
Our first goal is to meet an approximate version of the challenge
posed by Peng et al.  What we propose is that a site that wishes to
defend itself against DDoS attacks---we refer to such a site as a
\emph{defender}, denoting it usually by $D$---will characterize
explicitly the type of messages that it considers \emph{important};
important enough for $D$ to view itself as being well defended, if all the \emph{important}
messages sent to it during an attack would be delivered to it; and if
most of the non-\emph{important} messages are dropped, so that $D$ can
read the \emph{important} messages, and reply to them.  To
ensure that such messages would be delivered to $D$ under an attack,
we need an unequivocal, and easily recognizable, criterion for
discrimination
 between the messages that $D$ defined as \emph{important},  and all other
 messages sent to it. Such a criterion is introduced in the following section.

Our second goal is to use this criterion for constructing an effective
defenses against DDoS attacks.

\ss{The Proposed Criterion  of   Discrimination}\label{criterion}
To be effective for
defending against DDoS attacks, we need a criterion of discrimination
between the \emph{important} message for a given defender, and all
other messages sent to it.
Such criterion needs to satisfy the following two requirements.

\begin{enumerate}
\item This criterion should be easily recognizable, at the source of
      messages and at their destination. (The importance of
      recognizing the criterion at the source of messages---which is
      usually hard to do---will be  apparent in \secRef{DAD} where we
      introduce our second defense mechanism.)

\item  Messages classified as \emph{important} to $D$ by this criterion should not
be able to contribute meaningfully to a DDoS attack on it;
or, they should be trusted  by $D$  not to do so.
\end{enumerate}
\noindent
The ability to construct such a criterion depends on the manner in which
the set of \emph{important} messages for a given defender are characterized.

Note that the characterization of \emph{important} messages  cannot be based
on any predicate defined over their  content,
 because attackers
can learn about  such a predicate, and then send
messages characterized as important to  a given defender.
Moreover, the characterization of \emph{important} messages cannot be based
on the IP-address of the sender, because of the frequent use  of
IP-spoofing by attackers.

So, we propose that $D$ would characterize the messages that are
\emph{important} to it not by the structure of the message itself, nor by the
address of their sender, but
\emph{by  the behavior of their sender, before and
  during its process of sending messages to $D$}.
This behavior is to be defined by $D$ as a protocol of sending
messages---which we
 call \emph{a law} of $D$, denoting it by \law{D}.

 More specifically, $D$ would declare that
messages sent to it,  subject to the specified  law \law{D} are
\emph{important}.

For example,
 such a law may impose a  \emph{rate limiting} discipline on the
sender---say one message per 5 minute---which would prevent it to
contribute meaningfully to an attack on $D$.  Another possibility is
for a law to require that before starting to send messages to $D$, the
sender must authenticate itself in a specified manner.  This would
mean that a select group of senders, that $D$ presumably trusts, would
be able to communicate with $D$ during an attack on it.  And these two
measures can be combined into a single law.

But such  characterization of \emph{important}  messages
  may  seem to be  absurd---indeed,  how can anybody tell what is the
 law  that  governs the process of sending message by a given sender,
 if any?
Yet, this   is actually possible. It can be done, in particular,
 by using a middleware called \emph{law-governed
interaction} (LGI) \cite{min99-5,min05-8}.  Specifically, this
middleware features the concept of an \EL-message, which is a message
sent subject to a given law \EL, and which can be
 verified cryptographycally as such. We will discuss such messages in \secRef{L-message}.

\p{Summary:} The unequivocal criterion of
discrimination we propose is whether a message sent
to a given defender $D$ has been sent as an \EL-message, subject to the
law \law{D}  defined by $D$.

The manner in which such messages can be sent, and way they can be
recognized as \emph{important} messages with respect to a given
defender, are discussed in \secRef{L-message}.

Note that this criterion reverses the manner in which defense against
DDoS attacks is currently carried out. Conventional defenses operate by
trying to recognize attack-packets; and dropping them while letting
all other packets to pass through. On the other hand, the defense
based on our criterion of discrimination would recognize the messages
and packets that the defender declared to be important to it; and
letting them pass through, while dropping other messages and packets.

But although these are very different anti-DDoS approaches, they are
complementary and can be combined, as we do in both of the defense
mechanisms introduced in this paper.

 \ss{State-of-the-Art of Defense Against DDoS Attacks}\label{art}
Before introducing new anti-DDoS defense mechanisms, it is
appropriate  to review the state-of-the-art of such defense.
 
Practically all current defense mechanisms are based on techniques of identifying
\emph{attack-packets}, in order to drop them before they arrive at the
attacked defender.  The identification of
attack-packets
 is based on an analysis of many past DDoS
attacks---using statistical techniques and machine learning---coming
up with what is called a
 \emph{signature} (or signatures) of attack-packets---defined, mostly
over the content of packets.

This approach  gave rise, recently, to
 defense by means of \emph{scrubbing centers} \cite{ost20-1},
 operating by commercial companies such as Akamai and Cloudflare.
 These centers operate, essentially, by intercepting the messages
 addressed to a given defender $D$---their client---carrying out
 \emph{deep packet inspection} (DPI), of the packets addressed to the
 site they defend, in an attempt to identify and drop attack packets.

\sss{On the Advantages and Drawbacks of Scrubbing
Centers}\label{drawback}
 Scrubbing centers, or scrubbers for short, provide a decent, also flawed, defense
 against DDoS attacks. And  perhaps their biggest advantage is that they can be deployed
 practically anywhere over the Internet, since they do not depend on
 any special support of  routers.  But scrubbers
also have several substantial drawbacks, three of which are described  below.
\begin{enumerate}
\item \textbf{Unequal arm race:}  The ability of scrubbing to defend an attacked site is  undermined
 when confronted with a new kind of attack---because scrubbing is 
based on the analysis of past attacks, assuming that the current
attack resembles them, which it may not do completely. This creates an
enduring ``arm race'' between the  attackers and the defenders.  And it
is an unequal arm race, as it is far easier  and cheaper to mount a new type
of DDoS attack, than it is to incorporate the new attack into a new versions of the
scrubber. Thus, to quote from   \cite{ost20-1},
``we have been losing ground to
our adversaries in the DDoS war, and
we must take corrective action now''.

\item  \textbf{Unreliability:} The underlying purpose of scrubbing is to enable a defender
 to receive ``legitimate'' messages and to respond to them. But the
 ability of scrubbers to satisfy this purpose is unreliable due to the
 lack of definition of which messages are legitimate with respect of a
 given defender, and because the process of scrubbing may drop
many  legitimate messages and packets.

\item \textbf{High Cost:} Scrubbing center are expensive to operate
  and thus expensive to be defended by.
For example, according to information we recently obtained, Akamai charges
   \$1500 a month for the basic service, and a lot more for more
   advanced ones. CloudFlare
 provides free protection for very small sites,
    like personal websites, but the price goes up  steeply for 
    larger sites.  And without an expensive long term subscription one is left
    basically    defenseless. Although some  scrubbing centers
provide emergency on-boarding during an attack; such on-boarding takes time
while the client remain defenseless, and may carry  a hefty fee.
\end{enumerate}

\ss{On  the Use of our Discrimination Criterion for Defending Against DDoS
  Attacks}\label{dual-defense}

We offer two defense mechanisms, based on our  discrimination
criterion. 
The first mechanism is an upgrade of the conventional concept of
scrubbing center, making it more reliable by delivering to a defender
the  messages it defined as \emph{important}.
We characterize this mechanism as a \emph{discriminating
scrubbing mechanisms}, calling it
\emph{scrubbing++} (with apologies to, or honor of, Bjarne
Stroustrup).

Our second mechanism is a brand new mechanism, which depend on a
lightweight support by routers. It is called \emph{Discriminating
Anti-DDoS}, or DAD for short

Although both of these mechanisms are based on our discrimination
criterion, and thus share some of their aspects, they are very different and
each of them has its pros and cons.

\p{The rest of this paper is organized as follows:}
\secRef{L-message}  introduces the concept of an \EL-Message, on which
our discrimination criterion depends;
\secRef{scrubbing} introduces  the  \emph{scrubbing++ mechanism};
\secRef{DAD} introduces the DAD mechanism;
\secRef{eval} is an evaluation of these two defense mechanism;
and \secRef{conc} concludes this paper.

\s{The Concept of an \EL-Message}\label{L-message}
\begin{definition}
  An \EL-message is a message sent subject to a law \EL\, such that any receiver
of this message can 
determine, with 
justified confidence, that it has been sent subject to this law.
\end{definition}
\noindent
As pointed out in \secRef{intro}, an \EL-message, governed by
law \law{D} chosen by a defender $D$, is what one should
send to  $D$ to be considered important by it---particularly during
an attack. Such messages addressed to $D$ are denoted by  \law{D}-messages.

The concept of \EL-message, as described here,  is a very basic part of the middleware called
\emph{Law-Governed Interaction} (LGI) \cite{min99-5}.
 Here we just explain how \EL-messages are sent under LGI, and how
 they are identified as such.

\textbf{A note:}  To  use the concept of
\EL-message  one does not need to be
familiar with the entire LGI mechanism, but only  to read the
rest of this section.

\ss{Sending \EL-messages}\label{send-L}
In order to send \EL-messages, a sender $s$ must employ
 a software entity called the \emph{private controller} of $s$,
 denoted by $T^{\mathcal{L}}_{x}$, to serve as its surrogate by mediating
 the interactions of sender $s$
with others, subject to the given law  \EL.

 Such a private controller can be generated by $s$ via the following two
 steps:
 First,  acquiring the use of a \emph{generic controller} $T$ from
a  trusted service, called a \emph{controller-server} (CoS), that
 maintains a large number of \emph{generic controllers}
 (cf. \secRef{CoS}). A generic controller is built to serve as a
 surrogate of any given message-sender or message-receiver, subject 
to any given LGI-laws---with no specific laws built into it.

 The second step is to  load a law \EL\ into this generic controller,
 thus forming its private controller  $T^{\mathcal{L}}_{x}$.
Once this is done, $s$ can start sending \EL-messages via this
controller.

Note, however, that a law \EL\ may refuse to serve the particular sender
$s$---more about which, later.

The above two-steps are basically what a  software process needs
to do to send \EL-messages. But tools can be build  for making it easier for a human sender to
send \EL-messages. In particular, for senders operating via a smart
phone an app can be developed to carry out this process. (We have done
that with an experimental smartphone, as described in \cite{min10-5}.)
And  a browser-extension can be built to
help sending \EL-messages via a browser.

\ss{Authenticating an \EL-Message:}\label{verify-L} First note that
every LGI-controller carries a certificate signed by the controller
server (CoS) (acting as a CA), which, in effect, vouches for the
controller's authenticity.  Now, the authentication is done by means of
a TLS handshake between the controller and the receiver of this
message\footnote{We assume here that the receiver does not
operate via an LGI-controller, if it does then a different kind of
handshake is used.}, During this handshake the receiver of the message
obtains the certificate from the controller, validates it to ensure
that it's signed by the CoS---of course, the validation of the
certificate signed by the CoS requires the
receiver to have the public key of the CoS. The receiver also  gets
the controller's law hash. Moreover, the session's symmetric encryption key is also established during the
TLS handshake. The key exchange is based on Diffie-Hellman kex
algorithm.

Note that the receiver gets the message itself
 only  if the handshake succeeds.

 \ss{A Controller Service (CoS)}\label{CoS}
 To  support a large number of senders of \EL-messages, which may operate under a
variety of laws, one needs to provide a large set of generic controllers that
can be widely trusted to operate in compliance with any
valid law loaded into them.
As pointed out above, we call such a provider of controllers
a \emph{controller service} (CoS).
Its function is to  create,  maintain, and 
certify  a   collection of controller.
The certification is done by   providing each controller with a certificate of its
authenticity, signed by the CoS. And the CoS may be
geographically distributed.

  The current implementation of LGI includes an experimental version of
a CoS. But the proposed defense against DDoS attacks would require a
large scale CoS that needs to be managed by a reputable commercial
company or governmental institution. This organization must be willing
to vouch for the trustworthiness of its controllers, and to assume
liability for their failures.

Finally, it should be pointed out that the controllers maintained  by
the CoS can be used by many other applications besides the defense
against DDoS, such as making
heterogeneous distributed systems dependable \cite{min18-1}; building
decentralized social networks \cite{min16-1};  etc.

\ss{On the Roles of Laws in the Proposed Defenses Against DDoS}\label{laws}
Besides providing an unambiguous criterion for discrimination between
 \law{D}-messages address to a defender $D$---the messages that $D$
 declared as \emph{important} for it---the
law \law{D} can provide  the following benefits:
\begin{itemize}
\item  Limiting the ability of those that can send to $D$ \emph{important}
       message to it, 
       to contribute meaningfully to an attack on $D$.

\item Ensuring that the senders that can send  to $D$ messages that are
      \emph{important} to it, belong
      to the type of senders that $D$ would want to communicate with,
      even under attack.
\item Refusing to serve as a surrogate of some potential senders.      
  \end{itemize}
\noindent
Here we will discuss, briefly, three types of measures that can be
implemented, either separately or together, by  a law
\law{D}, for providing these benefits.
(Two of them were mentioned, more briefly, in \secRef{intro}.)

It is worth pointing out, before we start, that although some of these
measures, like authentication of the sender, are computationally
quite expensive, this computation is to be done by the controller
and, perhaps, the sender, and not by the routers.

\p{1: Rate Control:} This is done by having a law \law{D} 
impose  an upper bound on the  rate of messages that \law{D}-agents
can send to $D$. This is a very simple measure, which can be
implemented in several ways. And note that we cannot control the rate of
messages sent by a sender to its  controller. But the controller can
limit the rate of messages sent to the defender $D$.

Another rate control method is the following: Initially, every sender $s$
operating under the given law \law{D} is authorized to send just one message
to $D$. But $D$ can  enable $x$ to send more messages to it by
sending an appropriate message  to the controller of $x$, specifying
the  rate that it allows it to send messages.

\p{2: Certification:}
A law may be written to require a sender that tries to adopt it to
authenticate itself via one or several certificates. For example, one
of these certificates may be provided by $D$ itself for its best
customers. And supposing that $D$ is a military site, another certificate may
authenticate the sender's security classification level.

\p{3: Refusing to Serve as a surrogate:}
A law that requires certification can be written to refuse to accept
the sender $s$ as its client if it  did not satisfy its certification requirements.

The law can also send a captcha puzzle to the sender, dropping this
sender if it fails to solve it.

\p{Comments About Usage:}
First, we expect that most users will send their \EL-messages to
defenders via tools developed in browsers and smartphone, as described above.

Second, we expect the rate-control to be used most frequently in the
\law{D} of most defenders. On the other hand, certification is likely
to be used in such laws only for employees of $D$, or for its key clients,
and peers. And note that a single law can  contain either of these
options, or all  of them.

Third, that a defender may define several laws 
for sending messages which it consider important, but for simplicity,
we discuss just a single such law per  defender.

Finally, a defender $D$ that have large numbers of clients and
employees, would like to curtail the number of \emph{important}
messages that can sent to it. This can be done by having its law
\law{D} require certification via a certificate that $D$ will provide
its most important clients, employees and  peers. This
besides imposing rate limits on all such senders.

\s{Scrubbing++: A Discriminating Scrubbing Mechanism}\label{scrubbing}
One of the  drawbacks of conventional scrubbing centers (or scrubber,
for short) mentioned in \secRef{art},
is their lack of reliability, in that they are  likely
        to drop some of the legitimate packets---i.e., packets
        that the defender would like to get.
(Conventional scrubbers cannot really address this issue , because
they have no way of knowing which packets a given defender likes to
get.)

But the reliability of conventional scrubbing centers can be enhanced 
by having such a center  adopt our \emph{criterion of
  discrimination}, and use it to form what we call scrubbing++ center,
or   \emph{scrubber++}, for short.

   \p{The Rest of this Section is Organized as Follows:}
\secRef{gist1} describes the gist of this mechanisms;
\secRef{infra-1} introduces the  infrastructure of $S$;
\secRef{preparation-1} discusses preparations of $S$ for its role as a defender;
\secRef{sending-1} discusses the process of  sending  \EL-messages
to $P$;
\secRef{task-1} describes the treatment of \EL-messages arriving at  $P$;
 \secRef{port-1} describes the defense of $P$ from DDoS attacks;
Finally, \secRef{summary1} is a summary of the defense provided by scrubbing++.

\ss{The Gist of the Scrubber++ Mechanism}\label{gist1}
    The purpose of this mechanism  is to ensure that practically all
    \law{D}-messages sent to $D$, which
are the types of messages that $D$ declared
    as important to it, would be delivered to it.

For a given Scrubbing center  to upgrade to a  \emph{scrubber++},
which we denote by $S$, it
needs to  add an application called \emph{appendix}, denoted by $P$,
which operates via its own port $p$ and handles \EL-messages. Of course the \EL-messages addressed
to the host scrubber (denoted by $SC$)  should direct their messages
to  port  $p$.
Moreover,  the scrubbing center $SC$ on which $S$ is based \textbf{needs to avoid
scrubbing messages addressed to $P$}, in order to prevent packets sent
to $P$ from being dropped.

The main task of the appendix---relative to a given defender
 $D$ under attack---is to verify that  the \EL-messages addressed to $D$
are  \emph{important} to it. Namely, that they are \law{D}-messages,  were \law{D} is the
    law defined by $D$ to characterize message important to it.
And messages that are recognized by $P$ as \law{D}-messages 
 are sent by it  to $D$, while the    others messages addressed to $D$
 are dropped.

\ss{The Infrastructure of the Scrubbing++ Mechanism}\label{infra-1}
Besides a scrubbing center $SC$, on which the
scrubbing++ mechanism $S$ is squarely based, $S$ requires the
existence of three additional components: (1)  the appendix, which was
 introduced briefly above; 
(2) a components called a \emph{registry}, which serves as the
administrative center for all the sites that are defended by a given
$S$; and  (3)
 the Controller Server (CoS), which was introduced in
\secRef{CoS}, but requires  additional discussion in this context.
In the following two paragraphs we introduce the registry, and discuss
the CoS. The  appendix is discussed later.

\p{The Registry:}
The registry, is the
administrative center of $S$.
Its role is to register  sites  to be defended  by
 $S$, and to supply the appendix with information about
 these cites. The functionality of the registry is discussed in detail
 later.

The registry may reside in the
$SC$ itself, or it may reside elsewhere,  and  it may serve
several different scrubbing++ mechanisms. (But for simplicity,  we
mostly assume in this paper
 that each registry administers only a single scrubbing++
 mechanisms, i.e., a single $S$.
 
 \p{The Controller Server (CoS):}
 The CoS was introduces in \secRef{CoS} as potentially serving many
 types of applications. But here we assume
 that it serves only the scrubbing++ mechanisms $S$. To play this
 role,  the appendix $P$ would send  the CoS its  port number $p$.
  And  the CoS should  supply this port to every controller being acquired by an
 actor, to enable it to send \EL-messages to $P$.
 
\sss{Defending the CoS and the Registry Against DDoS Attacks}\label{CoS1}
For the  CoS to be used by the senders of \EL-messages to $P$ it needs to be
defended against DDoS attacks.
But this cannot be done by means of $S$.
Because $S$ enables messages to get through only if they are
\EL-messages,  sent via a controller acquired from the CoS.
But one needs to send a regular message to the CoS to acquire a
controller.
And if the CoS is defended by $S$, the packets of such a message may
be dropped.
Therefore,  the CoS needs to be defended via one of the
conventional \emph{scrubbing centers}, such as $SC$ itself.

The registry, if it is not reside in our scrubber $SC$,  would also
need to be defended, and it can be defended by our $S$ or by $SC$.
The appendix $P$ would also needs  defense against DDoS, as discussed
in \secRef{port-1}.

\ss{Preparing for Defense}\label{preparation-1}
For a website $D$  to be defended by the Scrubbing++  Mechanism $S$, the following
steps need to be carried out: (1) defining a defense law \law{D};
(2) registering $D$ as a defender;
 (3) providing information to the registry; and (4) providing
     information to the appendix.

\p{1: Defining a Defense-Law \law{D}:}

\p{1: Defining a Defense-Law:}
It should be pointed out that a single defender can specify several
laws that define different types  of message-sending behaviors that it
consider \emph{important}. But for
the sake of simplicity, we will limit ourselves here to a single such
law per defender.

We distinguish between two ways for defining laws.
One way is for the defender to write its own law, or have somebody
else write a law for it. The other way is to adopt a law that somebody
else defined. The registry would have a list of laws that its
clients---the registered defenders---use.
 It is worth pointing out that the law that a given
 defender  $D$ is using as its \law{D} need  not be a secret, as the
 knowledge of it cannot really  help in attacking $D$.

\p{2: Registering $D$ as a defender:} 
Only registered sites would be defended by  the scrubber++.
So registration constitute a permission to be defended.
Such a permission, by the registry, is required for several reasons.

First, if anybody would be allowed to register, with any law, then would-be attackers can
do so, using laws that enables them to attack, such as laws that do
not require rate limitation. This may create  a \emph{reverse attack}
on $S$. So, the registry should be selective about giving such
a permission. It is also important for the registry to accept a given
site $D$ as a defender, only if its, law \law{D}, can be shown to
prevent actors operating under it to
contribute meaningfully to an attack on $D$

And second, registration would probably involve charging a fee for being a defender.
But we do not address here the nature or the size of such a fee.

\p{3: Providing Information to the Registry:}
A request by a site $D$ to be registered  as a defender should
include the following  two pieces of information: First,  the pair  [$D$,H(\law{D})]
 where $D$ is the IP-address of would-be a defender, and H(\law{D})
 is the one-way hash of its law \law{D}.

And second,  $D$ will submit to the registry the text of the law \law{D}.
This can be useful for two reasons: (a) anybody who wants to 
 send \law{D}-messages to $D$, can find in the registry the law \law{D},
 which they need to load
their controller with.
And (b) a  potential defender that does not know how to write LGI-laws can
find a suitable law among those submitted to the registry.

 \p{4: Providing information to the appendix:}
For each registered defender $D$, the registry would disseminate the
pair  [$D$, H(L($D$))] to the appendix.

It will also provide the appendix
with the public key of the CoS, to be uses for decrypting the certificates signed by the CoS,
and planted in all the controllers managed by it. We leave it
unspecified how this information should be provided to the appendix.

\ss{Sending \law{D}-Messages to a Defender $D$}\label{sending-1} The
act of sending an \EL-message was described in \secRef{L-message}, but
the sending of such a message under the scrubbing++ mechanism differs
as follows:
 Such a message should by addressed not to the scrubbing center $SC$
that serves as the basis for a scrubbing++ mechanism, but to its
appendix $P$, which operates via its own port $p$.

This change in the  sending of \EL-messages needs to be set up in the browser
 extension and in the sending app on the smart phone,
 respectively (cf. \secRef{L-message}).

\ss{The Treatment of Messages Arriving at  the Appendix}\label{task-1}
 First, recall that the messages that arrive at the appendix are not
 being scrubbed (cf. \secRef{gist1}).
Now, we distinguish between three types of messages that require
different kind of treatment by $P$. They are  discussed below:

\p{(1) Messages addressed to a site which is not a
registered defender:} Such messages would   be forwarded to its
destination  without any verification of
their nature. But they can also be dropped
 as they are likely to be 
attack messages.

\p{(2) Messages addressed to a defender $D$ which is  not under attack:}
Such messages would be  forwarded to $D$, without any verification of
their nature.   Because, we assume that a defender not
under attack is able to verify its own messages.

But if some packets are missing in a given message, $P$ might
wait, for a reasonable time, for the
 missing packets  to be  completed by the
sender.

\p{(3) Messages addressed a defender $D$ under attack:}
In this case,  the appendix $P$ should
verifying that $m$ is \emph{important} to $D$.
That is,  $P$ would verify that m was sent by a  genuine
LGI-controller operating
subject to the law \law{D}. We denote such an \EL-message by \law{D}-message.

If this verification succeeds then $m$ should be sent to $D$---perhaps
after waiting for a reasonable time, for 
 missing packets  to be  completed by the
sender. But if the verification fails, 
 $m$ should be dropped.
 
The verification has two steps.
First, as explained in \secRef{verify-L}, the 
 controller that operates as the surrogate of the sender of the
 \EL-message conducts a TLS-handshake with the receiver of
 this message--- $P$ in this case. During this handshake, $P$    validates
the controller's certificate to ensure it is signed by the CoS, and
gets the controller's  hash  $h$ of the law, under which the controller
operates. 

Second, as explained in \secRef{preparation-1}, the appendix $P$ has
in its possession the list of pairs
[$d$,H(\law{d})], one for every defender $d$.
Now, $P$ would verify that there is a pair [$D$, h] in this list,
which means that the controller in question operate subject to the law
that $D$ selected. 

\ss{Defending the Appendix from DDoS Attacks}\label{port-1}
  Attackers are likely to discover the existence of the appendix and
  its role. And if port $p$ is fixed for long enough time,  attackers
  would eventually
  discover it, and may  mount an attack on it. To prevent such an
  attack we will  make $p$ change dynamically, and in a manner that
  makes such changes hard to discover, or predict.
    The following is a  strategy to do  that.

 Following  starting port number $p$ of $P$, we let $P$
 change its port number regularly, but in randomized intervals, as follows:
    Consider a time $t1$ when $P$ started to operates under  port $p1$.
  We have  $P$ send to the CoS the following message at this time:
          \emph{(p1,t1+dt,p2)}, which means, essentially, the
          following: \emph{the current port number is p1, it would
          change at time t1+dt to p2}. In other words, both $P$ and
          the CoS would change the port number to $p2$ at time
    \emph{t1+dt}.

    The time period \emph{dt} should satisfy two conditions: (a) it
    should be smaller, say half, of the estimated time that it takes
    for attackers to mount a new attack with a new port number; (b)
    $dt$ should be randomized, for obvious reasons. This
    transformation of port numbers is to be repeated recursively.

\p{A Complicating Factor:} There is another issue to be considered:
When the change from port
    number  $p1$ to $p2$ is made, by both $P$ and the CoS, there would
    likely be several controllers that still operate with the port
    number $p1$. The problem is that the messages sent by these
    controller would not arrive at $P$, which moved to a different
    gate, and would thus be lost.

    This issue can be addressed by a pair of remedies.
    First, the CoS should update the gate numbers of the
    controllers still  addressing port $p1$,  to $p2$.
 But this would take some time, so we need to carry out the second
 remedy, described below.

     When $P$ changes its gate number to $p2$, let it leave the previous
    gate $p1$ to still function as $P$. In other words, we have two
    instances of $P$ operating at the same time. We call them
    $P_{previous}$, that operates via  gate $p1$,
    and $P_{current}$ that operates  via gate $p2$.
    So, the   messages sent    by controllers that
    still use gate $p1$, would arrive at $P_{previous}$. And this
    situation would be maintained invariant of the sequence of gate
    numbers selected by $P$. Moreover, the scrubbing center $SC$ on
    which $S$ is based
    \emph{would  avoid
scrubbing messages addressed to both these versions of $P$}

Note that although unlikely, the attackers may still manage to mount an attack,  perhaps on
$P_{previous}$, which has longer life span than  $P_{current}$, but such an attack can last
for only brief amounts of time, due to the relentless changes of the
port numbers under which $P$ operates.

\ss{The Defense: a Summary}\label{summary1}
The scrubbing++ mechanisms provides two complementary defenses to
registered defenders. One is the assurance to a defender $D$ that  it
will receive practically all the \law{D}-messages sent to it. The
other defense  carried out by underlying grabber $SC$
is the dropping of large number of attack-packets, which
should enable $D$ to read the \law{D}-messages it gets, and to
respond to them.

Notes that the scrubbing mechanism $SC$ is likely to drop a sufficient
number of packets to enable $D$ to handle the
messages that arrive at it. Some of these messages would be the \emph{important}
message sent to it by the appendix. But there may be other messages
that may arrive at $D$, because the scrubber did not scrub them away, and which 
 may interest $D$, although they do not belong to the set of
\emph{important} messages.

\s{A Router-Based Discriminating Anti-DDoS (DAD) Defense
Mechanism}\label{DAD} The big difference between this mechanism and
scrubbing++, is that DAD relies on the support of routers. Despite
this fundamental difference between these two mechanisms there are
considerable similarity between them due to the fact that both rely on
the same criterion of discrimination between messages the defender
declares as \emph{important} to it and all other messages.
 When we get to
such a similarity
we will either refer here to the relevant part in
 \secRef{scrubbing}, or we will 
 repeat here  verbatim a piece of text from
 \secRef{scrubbing}.

 \p{The Rest of this Section is organized as follows:} 
\secRef{routers} described the role of routers in this defense, and
describes the reasons for  routers to agree to adopt this role:
\secRef{gist-2} presents the Gist of this mechanism;
\secRef{support-2} describes the infrastructure of DAD;
\secRef{preparation-2} describes the preparation for defense;
\secRef{treatment} describes the treatment of \EL-messages;
\secRef{packets} describes the flow of a-packets and u-packets into and in routers;
Finally, \secRef{DAD1} describes the DAD's defense.

\ss{The Routers}\label{routers}
 We start with the recruitment of routers to support the operation of
 DAD, and we continue with the very lightweight requirements that DAD makes from the
 routers.
 
 \sss{Incremental Recruitment of Routers}\label{increment}
 Our main goal is to be able
to defend sites all over the 48 contiguous states of the US. This may
follow the implementation of DAD  in  the rest of North America, in
Hawaii, and perhaps in other places where one can trust the management
of routers.  But satisfying our main goal
 would require all the core routers in this vast region of the US to
 support DAD.
 This  is very  unlikely to
  happen in one fell swoop.
  What we need is to start incrementally,
  by having DAD operate in a smaller region.

A reasonable choice is the set of core router in one \emph{autonomous
system} (AS), managed by a given ISP---we will call such an
\emph{autonomous system} simply an \emph{ISP}, which is a much
better-known term than ``AS.''
And  this paper is presented in terms of a single such ISP,
assumed to satisfy the requirements of DAD.
We often refer to this ISP as ``our ISP,'' or simply ``ISP.''

Given such an ISP, DAD will be able to defend
 sites anywhere in  its domain, while the senders of message to such
 sites can reside anywhere on the Internet.

Our plan for incremental  recruitment is based on the following 
observation: 
 If the DAD mechanisms will be proven effective
 in a single ISP, it would be likely to create pressure on other ISPs to provide
 the required support for DAD. Because the
sites residing in the domains of other  ISPs are likely to exert
pressure on their ISPs,  to support DAD.
Moreover, these sites may offer a payment to their ISP to do that,
thus creating a direct financial incentive for the ISP to support DAD.
So,  one can expect that the various ISPs in the 48 contiguous states of
the US  would  incrementally   support DAD's requirements.
And DAD would be able operate on  any collection of ISPs which are pairwise
contiguous with each other.

\sss{The Requirements of DAD from the Routers that Support it:}\label{require}
We distinguish here between 
structural and behavioral requirements that DAD makes from the
routers of our ISP.  The reasons for these particular requirements will
become apparent in due course.

The main  structural requirement is
 for the header of every packet to
have one-bit field called the \emph{pass-field}. And we say that the
pass-field is \emph{on} if its value is 1, and \emph{off} when its value is 0. We
will see below what does the pass-field signify. Alternatively, the
pass-field may reside in the data-part of a packet, which should never
be encrypted. This, alternative can be used if adding a field to the
header of a packet would slow the router in a significant amount. We
assume here the first option.

The main behavioral requirement from the routers of our ISP is  \emph{not to accept} a packet with the pass-field on, unless it
is submitted to this router by its guard---more about which later.
Also, if asked to do so, the routers will drop all the packets with their
pass-field off addressed to a defender under attack.

There are a few 
 additional lightweight requirements, which will be introduced  in due course.

 \ss{the Gist of DAD}\label{gist-2}

The packets that flow in the routers of our ISP are
classified into two disjoint categories: (a) the packets addressed to a
defender $D$ that belong to messages that $D$ defines  as
\emph{important}, which we denote by \emph{i-packets}; and (b) all
other packets, which we denote by \emph{u-packets}.

 Under DAD,
i-packets are marked by having their pass field \emph{on} (their values is
1); and u-packets are marked by having their pass-field off (their
value is 0) (cf. \secRef{require}).
This makes it very easy and efficient to distinguish between these two
types of packets, which gives rise to the main defense technique by
DAD:

When a a defender $D$ is under attack, all u-packets addressed to
it---most of which are likely to be attack packets---are
\textbf{dropped at their source}. As we shall see, dropping such
packet at their source has important consequences, one of which is the
ability to discover the identity of the attackers that mastermind the
DDoS attack in question.

\ss{The Infrastructure of DAD}\label{support-2}
Besides the routers in our ISP, (cf. \secRef{routers}) DAD requires
the existence of three components:  the
controller service; the registry; and the guard. We elaborate on them,
briefly, below.

\sss{The Controller-Service (CoS), and its Defense}\label{CoS-2} The
CoS, introduced in \secRef{CoS}, needs to be
resident within the domain of our ISP.
And it  needs to be defended against DDoS attacks. But due to
consideration analogous to those in \secRef{CoS1} it cannot be
defended by DAD, which uses the CoS for its operation, so it must be
defender either  by a conventions scrabber, or by our scrubber++ mechanism.

\sss{The DAD-Registry}\label{registry-2}
The DAD-registry, or simply registry, is the
administrative center of  DAD and it need to reside
 within the domain of our ISP.
 The role of the registry in the DAD defense, and its functionality,
are discussed  in the following subsections.

Like the CoS, the registry needs to be protected against DDoS attacks.
But unlike the CoS the registry can be defended via DAD. It can also
be easily replicated---which will make it even more secure---because changes in the content of the registry
are relatively rare.

\sss{The Guard}
The guard is a device that  handles \EL-messages.
We use a Scrubbing++ center to serve as a guard--- with one
important modification to be described later.

We assume here that there is just one guard serving our ISP.
But in a large ISP, or in a system of routers consisting of routers
managed by several ISPs, we may have several guards.

The guard needs to be resident within the domain of our ISP, and it
need to be trusted by its edge router. To gain such trust the router
should authenticate the guard, probably via public key
cryptography. We will not elaborate here on such the authentication itself,
but we should address the following issue that it raises.

The authentication of a guard may be computationally expensive,
perhaps too expensive for a router to perform.
If this is the case the authentication can be done by
 autonomous component built into each router, which is running on
its own processor and is not involved in routing. This components of a
router, which we call
its \emph{supplement}, carries out the authentication of the guard, and
is not involved in the routing itself.

 \ss{Preparation for Defense}\label{preparation-2}
For a site $D$  to be defended by the DAD mechanism, the following
steps need to be carried out:
\begin{enumerate}
\item $D$ should define its \emph{defense
law}, i.e., the LGI-law
\law{D} that would serve as basis for identifying the messages
and packets that $D$ views as \emph{important}.

\item $D$ should Register in the DAD-registry as a defender.

\item $D$ should provide certain information to te  registry.

\item The registry should provide certain  information to the guard. 
\end{enumerate}
\noindent
The first three steps above, are identical to the three steps of
preparation of the scrubbing++ mechanism (cf. \secRef{preparation-1}),
while the fourth step is new here. We will not repeat them here, but
will spell out item number 4:

 \p{4: Providing Information to the Guard:}
For each registered defender $D$, the registry would disseminate the
pair  [$D$, H(L($D$))] to the guard. It will also provide the guard
with the public key of the CoS.

\ss{The Treatment of \EL-messages}\label{treatment}
\EL-messages are handled by a guard, implemented essentially as a
scrubbing++ mechanism (cf. \secRef{scrubbing}).
So, \EL-messages are treated by the guard
 almost exactly as they are
treated by the scrubbing++ mechanism, including the  verification that a given
\EL-message addressed to a defended $D$  under attack is the
\law{D}-message;
and the manner that the guard protect itself against DDoS attacks.

The only difference between the scrubbing++ and the guard, in their
treatment of \EL-messages, is that while the former forwards
\EL-messages to their destination without changing them, the latter
would change the individual packets of a message, before delivering them
to its edge-router. The way this is done is described below.

\ss{The flow of A-Packets and U-Packets Into Routers, and In Them}\label{packets}

\sss{The Ingress of I-Packets and U-Packets of \EL-Messages}

The guard carries out the three kind of treatments for three kinds of
\EL-messages 
described in
\secRef{task-1}.
But it handles the packets of these messages in the following manner:

(1) For a message $m$ addressed to a site which is not a
registered defender, the guard will turn the pass-field of each packet
of $m$ to
 be off (i.e., its value is set to 0), thus 
 making it into a u-packet.

 (2) For a message $m$  addressed to a defender $D$ that is  not under
 attack,  the guard will turn the pass-field of each packet of $m$ to
 be on (i.e., its value is set to 1), thus 
 making it into an i-packet.

 (3) For a  message $m$ addressed to a defender $D$ under attack,
     there are two cases to consider.  If the verification of this
     message as an \law{D}-message succeeds then the guard will
 turn  each  packet of $m$ to be an i-packet, as above.
 But if the verification fails, each packet of $m$ will turn to be a u-packet.

 In all these cases the transformed message $m$ is then 
 transferred to the edge-router of the guard, to be
routed to their destination.
 
\sss{The Ingress of Packets of Other Than \EL-Messages}
  Packets that belong to anything but \EL-messages, are, by
  definition,  u-packets. So, we need to make sure that such
packets have their  pass-field  zero (i.e., off).

 To Achieve this situation we
make the following requirement from routers
 (this is one of the two  main behavioral requirement  from the
 Routers under DAD, see \secRef{require}:
 \emph{every  router in our ISP would set the pass-field to zero  of every packet
submitted to it  either (a) directly from outside, excluding i-packets
submitted to the router by its guard; or (b) or from a router that
belongs to  another ISP, which does not operate
subject to DAD---this would require adding a new pass-field to each such packet.}

\p{An observation:} Normally, a packet sent from a given router
in a given ISP to its target in the same ISP ($D$ in our case) would
be routed via routers in the same ISP. But this may not always be the
case. So suppose that our packet was routed to an ISP that does not
follow the above mentioned rules of our routers,  would the pass-field
of our packet retain its value? 
 We think that it will, with a high probability,
because unless this router is rogue, which routers are generally not,
there is no reason for it to change the pass field, so it is very
likely to stay intact.

\sss{The Flow of I-Packets and U-Packets Through the Routers} 
Since DAD does not drop any packets---unless instructed to do so, as
we shall see below---there will be an uninterrupted  flow
of i-packets and u-packets to the various defender.

The ease of
discrimination between the two types of packets may be
useful in two cases:

It is useful for  defenders not under attack, as it makes it easy to
recognize packets that belong to \law{D}-messages, which the defender
may view as particularly important to focus on.

More importantly, it may also provide 
defenders under an attack a degree of resistance to it, by attempting
to drop all u-packets, and retaining only the i-packets.
But the attack maybe too strong for  a defender  to be able to handle
the i-messages formed from the retained i-packets.
 
In this case, the attacked defender can invoke a  stage of DAD,
called DAD1, to protect it.

\ss{The Defense}\label{DAD1}
DAD has a normally dormant stage, called DAD1, that can be invoked to
defend any given
registered defender $D$   under attack.
Once invoked,  DAD1 would instruct 
 all the routers  of our ISP to start dropping all
u-packets sent to $D$.

This dropping of u-packet  would be done at the their very  ingress to
the router. Such dropping of u-packets at their source would have two important consequences.
First, it would eliminate the 
  clogging of routers with u-packets, which would otherwise be allowed
  to flow over the system of routers.

  The second consequence of dropping u-packets at their source, may be
  even more important. It enable the identification of the members of
  the attacking botnet.
 And with careful analysis of the Internet
  traffic towards the members of the botnet in question,
  one should be able to identify the mastermind or masterminds  of the
  botnet---the real  attackers---and perhaps takes them down. This
  potential may prevent attacks on sites within our ISP altogether. Because
  would-be-attackers would be fearful of being discovered.

But this defense raises  two issues:
(1) who should invoke DAD1, and how; and (2) when should DAD1 be
    deactivated and by whom. We address these issues below.

\p{Who Should Invoke DAD1, and  How?}
The simple answer to this question is that  the attacked defender $D$ itself would invoke DAD1 after
noticing that it is under attack.
It should be able to notice an attack, because the large increase of
the nmber of u-packets it gets.

The  invocation of DAD1 can be done as follows:
$D$ will send  an appropriate command to the guard, which would send
that command to every router. And the router will start dropping u-packts at
their ingress.

We note that $D$ should be able to send such a message
to the guard, because its ability to recognize and drop the u-packed
coming at it should enable it to send messages. But even if $D$ cannot do this electronically, it can
advice the registry by phone about this attack, and the registry will
send the right command to the guard.

\p{When Should DAD1 be Deactivated, and How?}
First note that deactivation of DAD1  needs to be done as soon as
possible after the attack concludes,
because $D$ would not want  to lose the u-packets that may be of
interest to it and that it can  process when not under an attack.

But how can $D$ detect the conclusion of the attack on it?
After all, it does not get any u-packets since DAD1 was invoked, and it will not
see any, after the conclusion of the attack, until DAD1 is deactivated.
Moreover, the router close to $D$ are in a similar bind, because  they
would experience much lower congestion once DAD1 started to drop
packets addressed to $D$. 

One solution to this problem is that $D$ would deactivate DAD1, as a test, say
every hour. If it detects an attack it will invoke DAD1 again.
 And if there no attack, then deactivation is done.

  Another potential solution is for all the routers to report
  periodically to the guard
  the number of packets addressed to $D$ they handled.  The guard,
  in turn, would be able to conclude
 from these reports  if  the attack is over, and it can act
  appropriately.
  
The first solution above has the advantage  of not making new
  requirements from the routers.

 \s{Evaluation}\label{eval}
We have introduced two different defense mechanisms based on our
\emph{criterion of discrimination} between messages that a given
defender defined as \emph{impotent} to it, and all other message sent
to it.
One of thee mechanisms  is \emph{scrubbing++}, which is an upgrade of the
conventional concept of scrubbing center; and the other is a brand new
mechanism called \emph{DAD} which relies on the support of routers,and
which uses scrubbing++ as one of its tools. These mechanisms have
their different pros and cons, but they have this in common:

Both ensure the delivery of practically all messages addressed to a
defender $D$, which $D$ defined as \emph{important} to it. Therefore, they
both mitigate the unreliability drawback of the conventional scrubbing
centers (cf. \secRef{drawback}).

We will now spell out the very different pros and cons of each of
these mechanisms.

\p{The Pros and Cons of Scrubbing++:}
Like the conventional scrubbing    center, scrubbing++ can operate
practically anywhere over    the internet. This is a great advantage
over DAD.

But, like the conventional scrubbing centers it suffers    from
an  unequal arm race.
And it is expensive to operate, and thus expensive for defenders  to employ.

\p{The Pros and Cons of DAD:}
Unlike scrubbing++, DAD can operate only in a region where  it gets
the support of the routers. So it has a much narrower range of
applicability than scrubbing++.

On the other hand, where it can be used it has the following advantages:

First, it may suffer only
 marginally from  the unequal arm race with
attackers, as this arm race can be felt only in the operation of the guard.

second,      due  to the
dropping of u-packets, addressed to  defenders under attack, at their
source (cf. \secRef{DAD1}), DAD would eliminate the 
  clogging of routers with u-packets, which would otherwise be allowed
  to flow in the system of routers.

Third, the consequence of dropping u-packets at their source, may be
  even more important. As it enable the identification of the members of
  the attacking botnet, which should help in identifying its
  masterminds. This capability may end up freeing the region defended by DAD
  from DDoS attacks.

  And fourth, DAD provide a broad, relatively cheap, and enduring
  support for all
      registered sites resident  in the domain of our ISP, and there
      can be many of them. (The types
      of sites that are eligible for this support
      depends on the
      judgment of the registry.)

\p{Recommendation:}
The two defense mechanism introduced in this paper have distinct ranges
of applicability. Scrubber++ can be deployed practically everywhere over the
Internet.
And it is preferable over the conventional scrubbing center of which
it is a simple extension.

DAD can be used  only in a region where it can get the support of
routers.
And it is preferable over scrubbing++ in such a region, due to its
quite unique advantages. However, as we have seen, scrubber++ has a
role to play along with DAD, as it serves as one of its components.

\s{Conclusion}\label{conc}

This paper introduces an unequivocal criterion of
discrimination between two kind of messages and packets; those that
are defined as \emph{important} by a given defender, and those that are not.
This criterion gave rise to two novel, and very different defense mechanisms.

One of them,  called \emph{scrubber++}, which is a simple upgrade of the
conventional scrubbing centers, and eliminates some of its drawbacks.
This mechanism can be deployed practically everywhere over the Internet.

The other mechanism, called DAD,
requires lightweight support by router.
It has, thus, a limited applicability. But when DAD gets the support of
the routers in a given region, it provides powerful capabilities. One of them is that it
enables the identification of the attackers.

\bibliography{biblio}
\end{document}